\documentstyle[12pt]{article}
\begin{document}
\begin{center}

{\bf ABSENCE OF SINGULARITIES IN A COSMOLOGICAL PERFECT-FLUID SOLUTION\\}
 \vspace{1.5cm}
\small{F.J.CHINEA} and \small{L.FERN\'ANDEZ-JAMBRINA}\\
{\it Departamento de F\'{\i}sica Te\'orica II, Facultad de
Ciencias F\'{\i}sicas \\Ciudad Universitaria, 28040
Madrid, Spain\\ and\\Instituto de F\'{\i}sica Fundamental\\Ciudad
Universitaria,
 28040Madrid, Spain}\\
\vspace{5mm}
and\vspace{5mm}\\
\small{J.M.M.SENOVILLA}\\
{\it Departament de F\'{\i}sica Fonamental,Universitat de
Barcelona\\Diagonal 647, 08028 Barcelona, Spain\\and\\ Laboratori de
F\'{\i}sica Matem\`atica\\Societat Catalana de F\'{\i}sica, I.E.C.,
Barcelona, Spain\\} 
 \vspace{1cm}
\small{ABSTRACT}\\  

 \baselineskip 12pt
\parbox{30pc}{\vspace{2mm}\footnotesize{\hspace{.7pc} In this lecture we will
show some properties of a singularity-free solution to Einstein's equations and
its accordance with some theorems dealing with singularities. We will also
discuss the implications of the results.}}
\end{center}

\baselineskip 14pt
\vspace{1cm}
\noindent{\bf 1.Introduction}
\vspace{.5cm}

The occurrence of singularities in solutions to
Einstein's equations has been largely studied in the literature (see, for
instance,[1] and references therein).Recently, one of us presented a
cosmological, perfect-fluid solution whose curvature scalar polynomials were
finite [2] and this raised the question whether there was a singularity of other
kind.It has already been proven [3] that this space-time is singularity-free.
So, in this lecture, we would like to give an account of the properties that
follow from this fact. More details can be found in [3].

In section 2 we deal with the geodesic completeness of the solution,
whereas in section 3 we show some of its nice properties which are useful to
see how it agrees with the singularity theorems.This we will show in section
4.As there are many singularity theorems in the literature, we shall only
consider two of them :We shall deal with that of Penrose[1,6] and the most
powerful one, proven by Hawking and Ellis [1,7]. Section 5 is devoted to a brief
discussion of the results.
 
 \newpage \noindent{\bf 2.Geodesic completeness}
 \vspace{.5cm}

The solution presented in [2] describes a cosmological space-time
corresponding to a radiation-dominated universe exhibiting
cylindrical symmetry:
\begin{eqnarray}
ds^2&=&cosh^4(at)cosh^2(3ar)(-dt^2+dr^2)+\nonumber\\
 & &(9a^2)^{-1}cosh^4(at)sinh^2(3ar)cosh^{-2/3}(3ar)d\phi ^2+\nonumber\\
 & &cosh^{-2}(at)cosh^{-2/3}(3ar)dz^2
\end{eqnarray}
The density and pressure of the fluid take the following form:
\begin{eqnarray}
\chi\rho&=&15a^2cosh^{-4}(at)cosh^{-4}(3ar)
\end{eqnarray}
\begin{eqnarray}
p&=&\rho/3
\end{eqnarray}
where $\chi $ is the gravitational constant.

Using standard techniques, we get the following equations for the geodesic
motion: 
\begin{eqnarray}
\ddot{t}+2a~tanh(at)(\dot{t}^2+\dot{r}^2)+6a~tanh(3ar)\dot{t}\dot{r}+\nonumber\\
(2/9a)tanh(at)sh^2(3ar)cosh^{-8/3}(3ar)\dot{\phi}^2-\nonumber\\
a~cosh^{-7}(at)sinh(at)cosh^{-8/3}(3ar)\dot{z}^2&=&0
\end{eqnarray}
\begin{eqnarray}
\ddot{r}+3a~tanh(3ar)(\dot{t}^2+\dot{r}^2)+4a~tanh(at)\dot{t}\dot{r}-\nonumber\\
(1/9a)sinh(3ar)cosh^{-5/3}(3ar)[3-tanh^2(3ar)]\dot{\phi}^2+\nonumber\\
 a~cosh^{-6}(at)sinh(3ar)cosh^{-11/3}(3ar)\dot{z}^2&=&0 \end{eqnarray}
\begin{eqnarray} (9a)^{-1}cosh^4(at)sinh^2(3ar)cosh^{-2/3}(3ar)\dot{\phi}&=&K
\end{eqnarray}
\begin{eqnarray}
cosh^{-2}(at)cosh^{-2/3}(3ar)\dot{z}&=&L
\end{eqnarray}
\begin{eqnarray}
cosh^4(at)cosh^2(3ar)(-\dot{t}^2+\dot{r}^2)+\nonumber\\
 (9a^2)^{-1}cosh^4(at)sinh^2(3ar)cosh^{-2/3}(3ar)\dot{\phi}^2+\nonumber\\
 cosh^{-2}(at)cosh^{-2/3}(3ar)\dot{z}^2&=&-\delta
\end{eqnarray}
Here a dot means derivative with respect to the affine parameter, $K$ and $L$ are
constants of motion along the geodesics and $\delta $ takes the value one or zero
for timelike or null geodesics respectively. As all functions
involved are non-singular,the solutions exist and are unique.

It can be easily seen [3] that the geodesics do not grow faster than their
tangents for positive time coordinate in the vicinity of the axis (for small
enough radius) or for large $r$ or $t$.So the geodesics can be extended to
arbitrarily large values of the affine parameter.

Two families will be particularly interesting for the discussion: One is formed
by the congruences of ingoing and outgoing radial null geodesics through every
point of the manifold, whose motion is unbounded and its {\it radial speed} never
changes sign.The other one comprises the null geodesics on the hyperplanes
$z=const.$\/The geodesic motion along the latter is bounded in the radial
coordinate [3].

\vspace{1cm}
 \noindent{\bf 3.Properties of the solution}

\vspace{.5cm}
\noindent{\it 1.1.Global hyperbolicity}
\vspace{.2cm}

 It follows from section 2 that every maximally extended null geodesic meets any
of the hypersurfaces $t=const.$\/This means [4] that every non-spacelike curve
intersects the mentioned hypersurfaces only once (they are global Cauchy
surfaces) and so the solution is globally hyperbolic and fulfills the whole
hierarchy of causality conditions under it [5], for instance the chronology
condition (there are no closed timelike curves), which will be needed in the
next section.

\vspace{.5cm}
\noindent{\it 1.2.Singularity-free}
\vspace{.2cm}

 Since the $t$ is a cosmic time and the hypersurfaces
$t=const$ are Cauchy surfaces, every non-spacelike curve can be extended to
arbitrarily large values of the affine parameter. This means that the
solution is bundle-complete [1], which is the usual definition for lack of
singularities.

\vspace{.5cm}
\noindent{\it 1.3.Strong energy and generic conditions}
\vspace{.2cm}

 From the formulae (2,3) it is obvious that the energy-momentum tensor never
vanishes and $R_{ab}v^av^b>0$, so both conditions are satisfied [1,5].

\vspace{1cm}
 \noindent{\bf 4.Accordance with the singularity theorems}

\vspace{.5cm}
\noindent{\it 4.1 Penrose's theorem}
\vspace{.2cm}

 We have already proven that the solution is null geodesically complete,
globally hyperbolic and that the strong energy condition holds. Obviously, if it
is to be in accordance with the theorem [1,6], it must have no closed trapped
surface (a spacelike surface in which the traces of the two null second
fundamental forms have the same sign). Computing both traces in the point where
the radial coordinate reaches its maximum (it has got one since the surface is
compact), we get that they have opposite signs, so there is no closed trapped
surface.

\newpage
\noindent{\it 4.1 Hawking and Penrose's theorem}
\vspace{.2cm}

Since the space-time is geodesically complete and every other condition of the
theorem holds (generic, strong energy and chronology conditions), the whole set
of three alternative conditions must not be fulfilled[1,7]. We know that
there's no trapped surface, so we only have to check that the other two do not
hold:

1.Existence of a point $q$ such that on {\it every\/} past (future) null geodesic
from $q$ the expansion becomes negative: We can see such point does not exist
in this solution just remembering the two families of geodesics pointed out in
section 2. The radial null geodesics through any point in the manifold diverge
towards the future (past) if they are outgoing (ingoing), so their expansion
cannot be negative. Another way to see that the point $q$ does not exist is to
remember that through any $q$ there are null geodesics with $z=const.$ which are
bounded above and below in $r$. Thus, these geodesics can never converge with
the radial ones, which are unbounded in $r$.

2.The existence of a compact achronal set without edge: Suppose there is one.
Take  a point $q$ in the set. By using the radial geodesics we can always choose
points $q_-\in I^-(q)$ (the chronological past of $q$) and $q_+\in I^+(q)$ 
(the chronological future of $q$) such that $r(q_-)=r(q_+)>r(q)$, where for any
point $s$ we denote by $r(s)$ the value of the coordinate $r$ at $s$. Since
$q_+\in I^+(q_-)$ and $r(q_+)=r(q_-)$, we can join $q_-$ and $q_+$ with a
future-directed worldline of the fluid congruence. As the required achronal set
has no edge, this worldline intersects the set, and it will do it at a point
$\tilde{q}$ with $r(\tilde{q})=r(q_-)=r(q_+)>r(q)$. This proves that the coordinate $r$
 cannot be bounded for any achronal set without edge. It is obvious then
that any achronal set in the manifold cannot be both compact and without edge.

\vspace{1cm}
 \noindent{\bf 5.Discussion}
\vspace{.5cm}

We have shown that the solution [2] is singularity-free and in agreement with
some of the main singularity theorems. This illustrates the importance of the
initial conditions (existence of a trapped surface or an achronal set without
boundary) for the development of singularities, since the energy and causality
conditions are not determinant for their appearance.
There is no reason then why there should not be singularity-free solutions with
little or even no symmetry at all if they do not have the initial conditions
required by the singularity theorems. 

 \vspace{1cm}
 \noindent{\bf 6.Acknowledgements}
\vspace{.5cm}

The present work has been supported in part by DGICYT Project PB89-0142 (F.J.C.)
and CICYT Project AEN90-0061 (J.M.M.S); L.F.J. is supported by a FPI Predoctoral
Scholarship from Ministerio de Educaci\'{o}n y Ciencia (Spain). J.M.M.S. wishes to
thank E. Ruiz for discussions.

\newpage
 \noindent{\bf 7.References}
\vspace{.5cm}

\noindent 1. S.W. Hawking and G.F.R. Ellis, {\it The large scale
structure of space-time\/}, Cambridge Univ. Press, Cambridge
(1973).\\
2. J.M.M.Senovilla,{\it Phys. Rev. Lett.}\/{\bf 64}, 2219
(1990).\\
 3.F.J.Chinea, L.Fern\'andez-Jambrina and J.M.M. Senovilla, {\it 
 Phys. Rev. D} {\bf 45}, 481 (1992) [arXiv: 
 gr-qc/0403075]\\
4. R. Geroch, J. Math. Phys. {\bf 11}, 437 (1970).\\
5. J. Beem and P. Ehrlich, {\it Global Lorentzian Geometry\/},Dekker, New York
(1981).
6. R. Penrose, Phys. Rev. Lett. {\bf 14}, 57 (1965).\\
7. S.W. Hawking and R. Penrose, Proc. Roy. Soc. London A{\bf 314}, 529 (1970).

  \end{document}